\documentclass[twocolumn]{aastex63}
\usepackage{amsmath}
\usepackage{gensymb}
\usepackage{float}
\usepackage{booktabs}
\usepackage[title, titletoc]{appendix} 
\usepackage{xcolor}

\usepackage{verbatim}
\accepted{December 21, 2025}

\usepackage{lineno}

\graphicspath{{./}}

\begin{document}

\title{Disentangling the hemispheres of Teegarden's Star b with LIFE}

\correspondingauthor{Ryan Boukrouche}
\email{ryan.boukrouche@astro.su.se}

\author[0000-0002-5728-5129]{Ryan Boukrouche}
\affiliation{Department of Meteorology, Stockholm University, Sweden}
\affiliation{Department of Astronomy, Stockholm University, Sweden}

\author[0000-0001-8345-593X]{Markus Janson}
\affiliation{Department of Astronomy, Stockholm University, Sweden}

\begin{abstract}

Teegarden’s Star is one of the most promising targets for the first observations of LIFE, as a non-transiting rocky planet with similar bulk properties to the Earth, and a relatively quiescent M-dwarf host star. We use LIFEsim, a software developed by the ETH LIFE team, along with thermal emission maps obtained from a suite of three-dimensional global climate model (GCM) simulations, to explore the sensitivity of LIFE to the observation geometry. We find that 3 days of observation in broadband would be enough to disentangle the hemispheres of the planet with a 1$\sigma$ or 3$\sigma$ confidence level with a baseline or optimistic scenario respectively. Doing the same for a fast-rotator in the habitable zone of a G-class star would be prohibitively challenging. Given enough observation time, the sensitivity of LIFE may allow some spatial resolution of Teegarden's Star b to be achieved, which may directly link to the presence of water clouds and therefore an active hydrology.

\end{abstract}

\keywords{extrasolar rocky planets --- astrobiology --- habitable planets}

\section{Introduction} \label{sec:intro}

The large majority of the currently confirmed exoplanets transit in front of their stars. On the other hand, transit probabilities tell us that more than 95\% of all exoplanets do not transit. The task of increasing the number of non-transiting terrestrial exoplanets on our catalog is largely being taken on by facilities capable of radial velocity measurements whose precision might get close to the 9 cm/s required to detect an Earth twin \citep{hara2023statistical}. EXPRES (R=150,000) is expected to reach about 30 cm/s \citep{blackman2020performance}. One of the drawbacks of the radial velocity method is that the dependence on orbital inclination makes Doppler signals fainter for more face-on systems. Astrometry can bypass these considerations and could contribute significantly to the discovery of small non-transiting planets. An example of an astrometric mission conceived for this purpose is the Theia micro-arsecond astrometric observatory \citep{malbet2024high}. Astrometry does not depend on orbital parameters or the activity of the star, and would provide the exact mass and all the orbital parameters of terrestrial planets around nearby stars. 

The resulting list of terrestrial planets in their habitable zones, however long it may be, will be characterized by direct imaging observatories, particularly LIFE \citep{quanz2022large}, HWO \cite{vaughan2023chasing}, and ELT-PCS \citep{kasper2021pcs}. LIFE is preparing for two scenarios: either a sufficient number of targets are discovered by the time it launches, in which case the mission can immediately enter its characterization phase, or it needs a larger number of targets, in which case it will start with a search phase. 

These upcoming capabilities highlight the need for precursor science that will provide a better understanding of the observables we should expect for planets in various orbital configurations, astrophysical environments, and evolutionary stages.   

This study is a continuation of \cite{boukrouche2024impact} and \cite{boukrouche2025near} which respectively explored the potential of LIFE to disentangle different water cloud cover fractions and the habitability of Teegarden's Star b. Here we explore the sensitivity of LIFE to the orbital phase and inclination of the planet.

In Sect. \ref{sec:methods} we describe the tools used for the modeling and the parameters used. We present results in Sect. \ref{sec:results}. We discuss the significance of the results for future observations in Sect. \ref{sec:discussion} and conclude in Sect. \ref{sec:conclusion}.

\section{Methods} \label{sec:methods}

We use the results of simulations obtained with Isca \citep{vallis2018isca}, a flexible framework used to model the global circulation of planetary atmospheres, developed and maintained at the University of Exeter. It has been used previously to simulate the climate and atmospheric dynamics of tidally-locked exoplanets \citep{2018ApJ...868..147P,2022ApJ...941..171L}. For this work, Isca was configured as an idealized aquaplanet general circulation model and coupled to the correlated-$k$ radiative transfer code \textsc{socrates} \citep{edwards1996socrates} and a simple diagnostic representation of water clouds (Simcloud;  \citealp{liu2020simcloud}).
 
These simulations divided the spectral range into 16 bands to allow for reasonable runtimes in 3D while retaining a good accuracy. They use a range of instellations, including the one estimated for Teegarden b from the measurements performed by \citep{dreizler2024teegarden}, 1481 Wm$^{-2}$. Two different surface albedos were used, 0.07 to represent a water-covered surface, and 0.3 which represents a land-dominated surface. The series of simulations also included experiments where clouds were turned off and ones where an Earth-like ozone layer was added. None of these attempts significantly changed the results of the present work. Therefore, all the subsequent results discussed below will use a surface albedo of 0.07 and exclude ozone. 

We model two cases. The first is Teegarden b orbiting its M-class star at a separation of 0.0259 AU. In this case we assume the planet to be tidally-locked with the orbital parameters of Teegarden b as measured by \cite{dreizler2024teegarden}, meaning that the rotational period is equal to the measured orbital period of 4.90634 Earth days. We use a BT-Settl photospheric stellar spectrum from the Spanish Virtual Observatory with an effective temperature of 3000 K and a log(g) of 5, representing Teegarden's Star. The second case is a theoretical exoplanet like Teegarden b which might be discovered in the future, orbiting a G-class star at a separation such that it receives the same stellar constant as Teegarden b. With the Sun's luminosity, we find this separation to be 0.9586 AU. This case assumes that the planet is not tidally locked and instead adopts the same rotation rate as the Earth. Both cases assume a present-day Earth-like atmospheric composition with 78.084\% N$_2$ and 20.947\% O$_2$ as major components, and 400 ppmv CO$_2$, 1 ppmv CH$_4$, 0.001 ppmv CO, and 0.03 ppmv H$_2$ as trace species. H$_2$O has an abundance determined by a balance between precipitation and evaporation, and it is advected by atmospheric circulation. The surface pressure is 1 bar.

A more complete description of the model and simulation parameters can be found in the appendix section of \cite{boukrouche2025near}. For the present study, we have extended these simulations by an additional month using 400 bands, including 326 bands from 285 nm to 800 nm and 74 bands from 800 nm to 10 mm, allowing us to better resolve the resulting emission spectra along with potential molecular signatures.

The outgoing longwave radiation from Isca were input into LIFEsim \citep{dannert2022large}, a software developed by the ETH LIFE team capable of producing synthetic observations with the resolving power, wavelength coverage, and integration time of the instrument as the main instrumental inputs. There are other physical inputs summarized in Table \ref{tab:1}. The instrument's theoretical baseline performance is also complemented by the option to select an optimistic or a pessimistic scenario. The scenarios correspond to different aperture sizes of the unit telescopes that make up the interferometer. The pessimistic, baseline and optimistic scenarios respectively correspond to 1, 2, and 3.5 m diameter mirrors \citep{quanz2022large}. 

\begin{table}[tbh]
\centering
\begin{tabular}{lll}
\toprule
\toprule
\multicolumn{2}{l}{Star distance} & 3.832 pc \\
\multicolumn{2}{l}{Planet radius} & 1.02 R$_{\mathrm{Earth}}$ \\
\midrule
         & M star   & G star  \\
\midrule
Star radius     & 0.12 R$_{\mathrm{Sun}}$    & R$_{\mathrm{Sun}}$ \\
Star effective temperature   & 3034 K & 5772 K \\
Semi-major axis & 0.0259 AU & 0.9586 AU \\ 
Angular separation & 0.00676 asec & 0.25 asec \\
\bottomrule
\end{tabular}
\caption{Parameters used in LIFEsim.}
\label{tab:1}
\end{table}

Since the orbital geometry of Teegarden's Star b is not yet known, we explore six possible hemispheres which could be observed along the orbit. LIFEsim allows the computation of the statistical significance $\sigma$ of the detected difference between a reference hemisphere, here taken as the dayside centered on the substellar point, and the other hemispheres. It is computed as 
\begin{equation}
\sigma=\frac{|F_{\mathrm{day}}-F_{X}|}{\frac{F_{\mathrm{day}}}{\mathrm{SNR}_{\mathrm{day}}}},
\end{equation}
where $F_{\mathrm{day}}$ is the dayside flux, $F_{X}$ is the flux from the other hemispheres, and $\mathrm{SNR}_{\mathrm{day}}$ is the signal to noise ratio associated to the dayside flux observation.
 
\section{Results} \label{sec:results}

\begin{figure}[htb]
\centering
\includegraphics[width=0.49\textwidth]{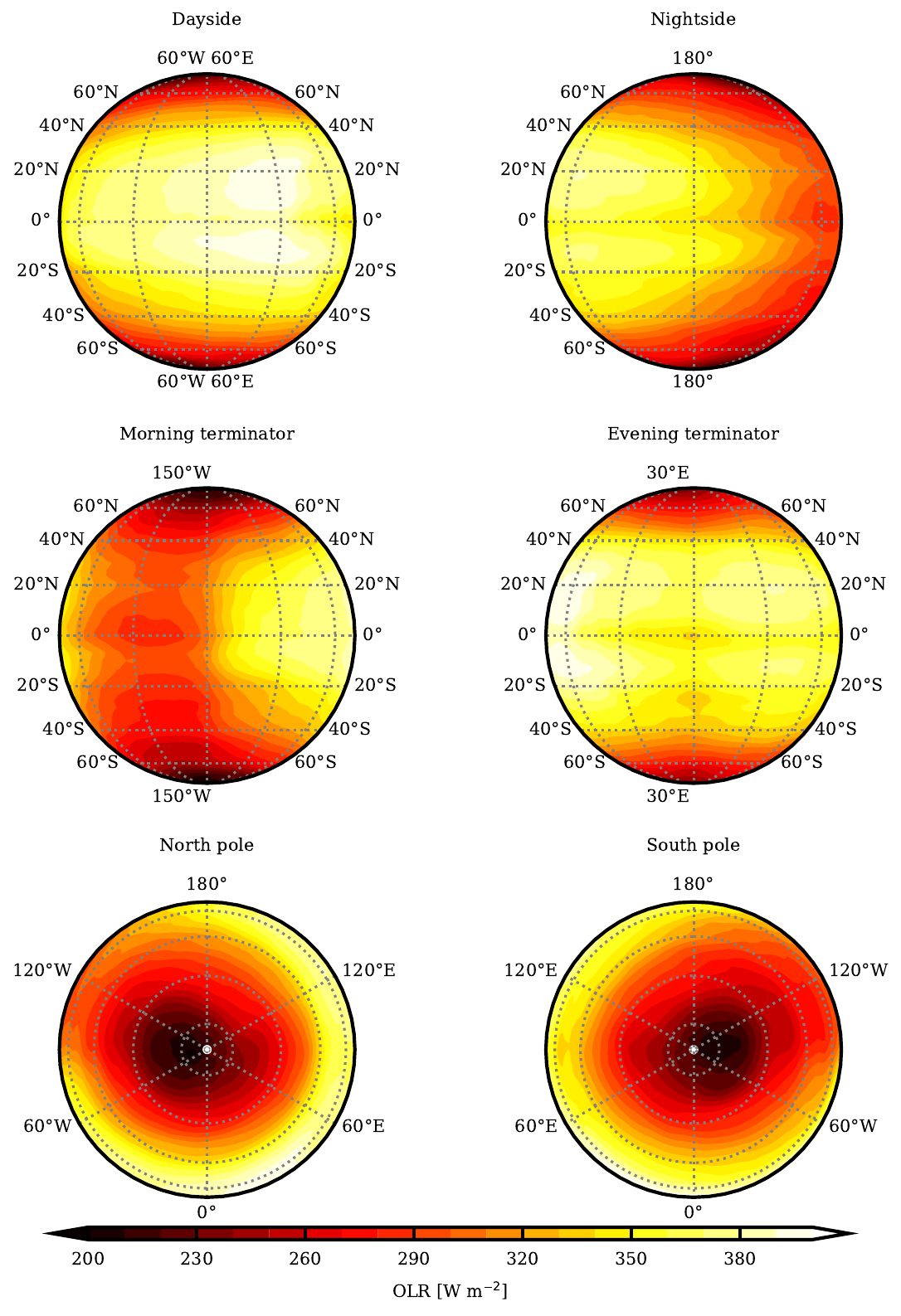} 
\caption{Outgoing longwave radiation of six potentially visible hemispheres of the planet.}
\label{fig:Mstar_Faces_soc_olr}
\end{figure}

%
\begin{figure}[htb]
\centering
\includegraphics[width=0.49\textwidth]{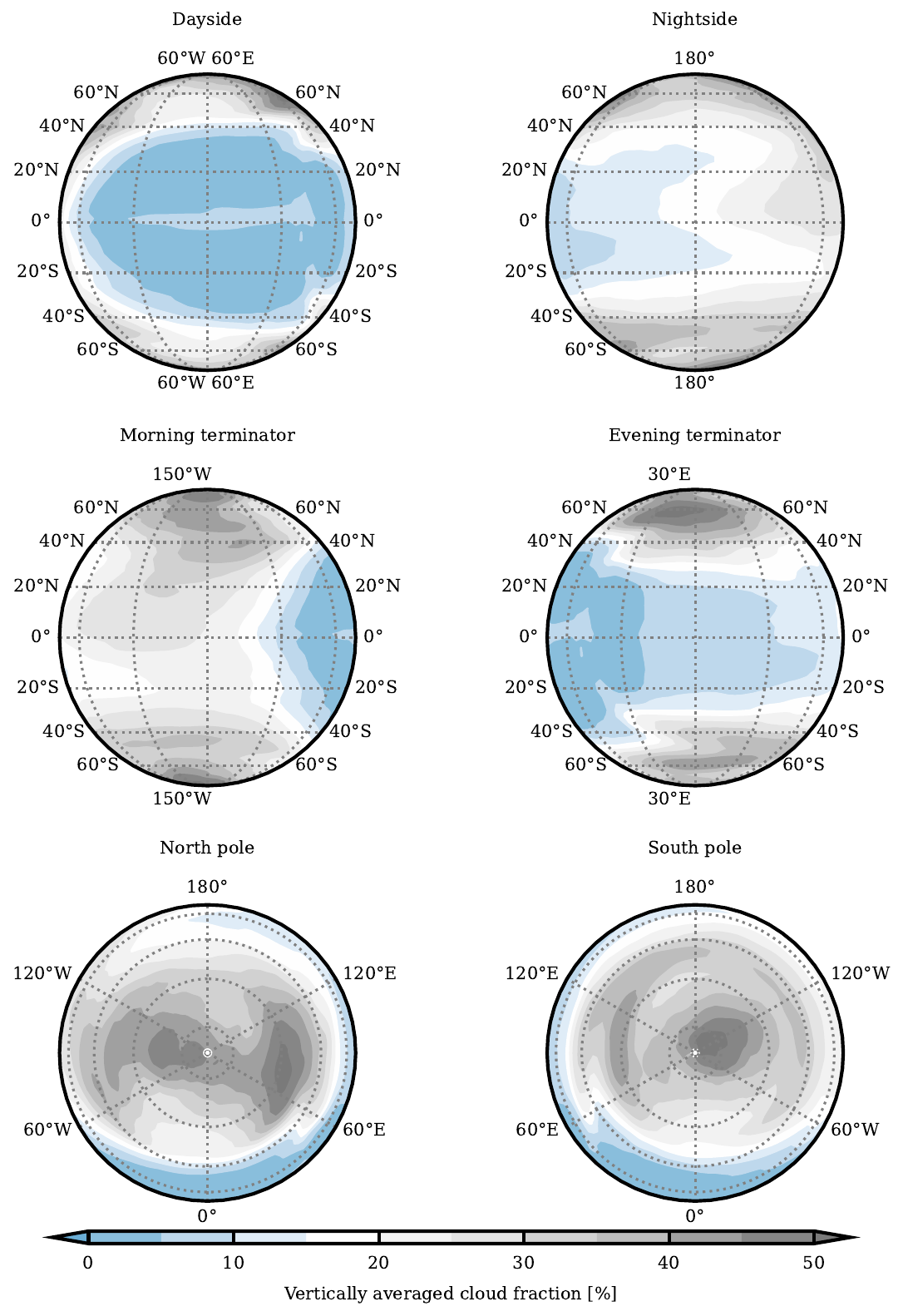} 
\caption{Cloud fraction averaged over all pressure levels.}
\label{fig:Mstar_Faces_cloud_avg}
\end{figure}
Figures \ref{fig:Mstar_Faces_soc_olr} and \ref{fig:Mstar_Faces_cloud_avg} show the thermal emission and the vertically averaged cloud fraction of Teegarden b seen from six different hemispheres, which is output from the simulations. The dayside does feature a cloud cover, however it is a high cloud deck around 1-10 mbar, which would be located in the stratosphere on Earth. It is optically thin enough that the planetary albedo is very small on this side, and this maximizes the resulting outgoing longwave radiation. The latter's pattern is mostly dominated by the cloud cover, and would be instead driven by specific humidity without clouds.

\begin{figure}[htb]
\centering
\includegraphics[width=0.49\textwidth]{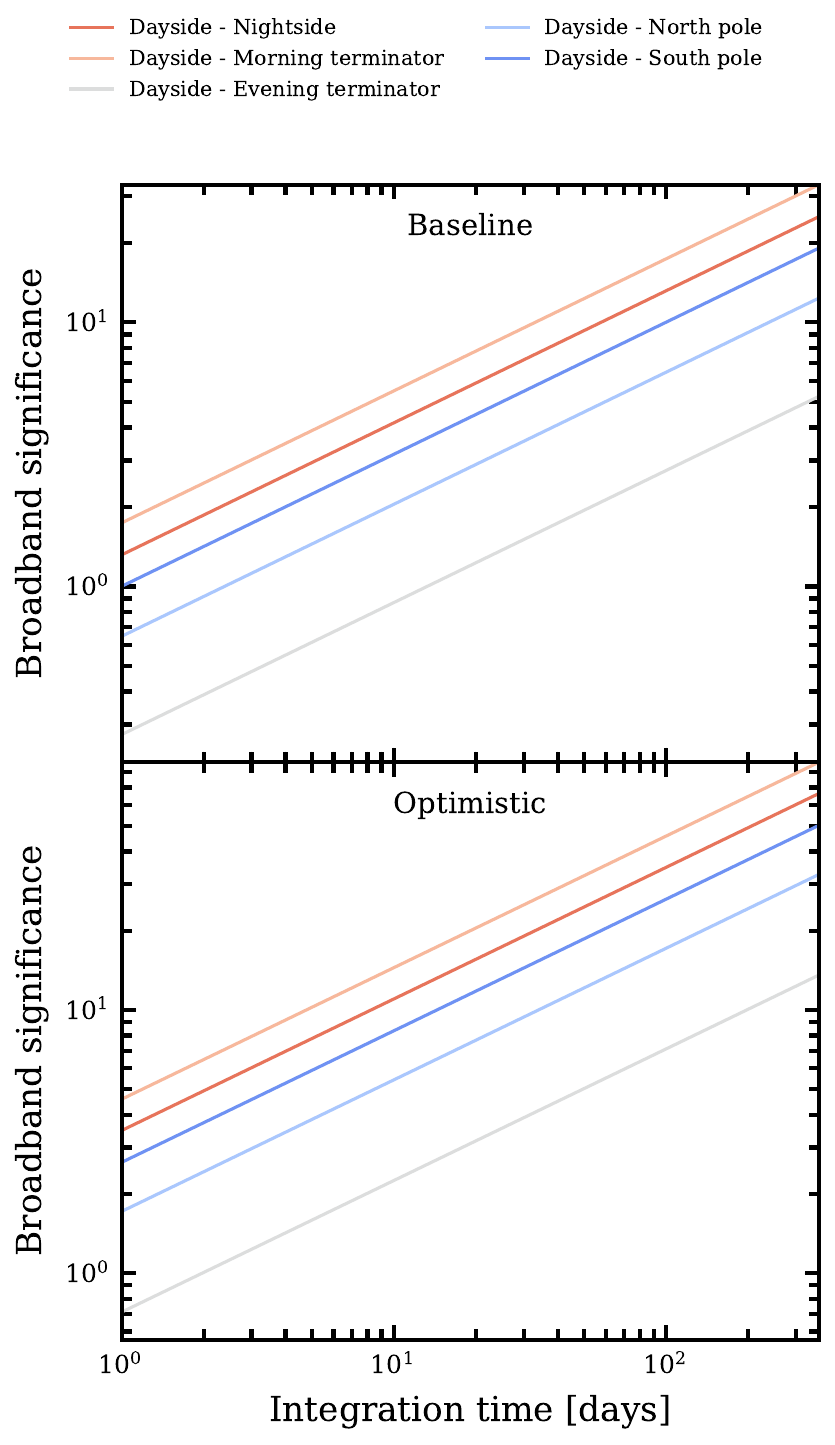} 
\caption{Broadband significance as a function of integration time for Teegarden's Star b, assuming an instellation of 1481 Wm$^{-2}$ and either a baseline or an optimistic scenario. The dayside hemisphere is taken as the reference.}
\label{fig:broadband_Mstar}
\end{figure}

Figure \ref{fig:broadband_Mstar} shows the broadband statistical significance of the detected difference between the dayside and each of the other hemispheres, computed as $\sigma_{bb} = \sqrt{\sum_i\sigma_i^2}$, where $i$ is the index of the spectral bin. It shows that, as far as broadband properties are concerned, confidence levels higher than 1$\sigma$ can be achieved in most cases with 3 days of observation and a baseline scenario, and as high as 3$\sigma$ in the same time with the optimistic scenario. 

\begin{figure*}[htb]
\centering
\includegraphics[width=0.99\textwidth]{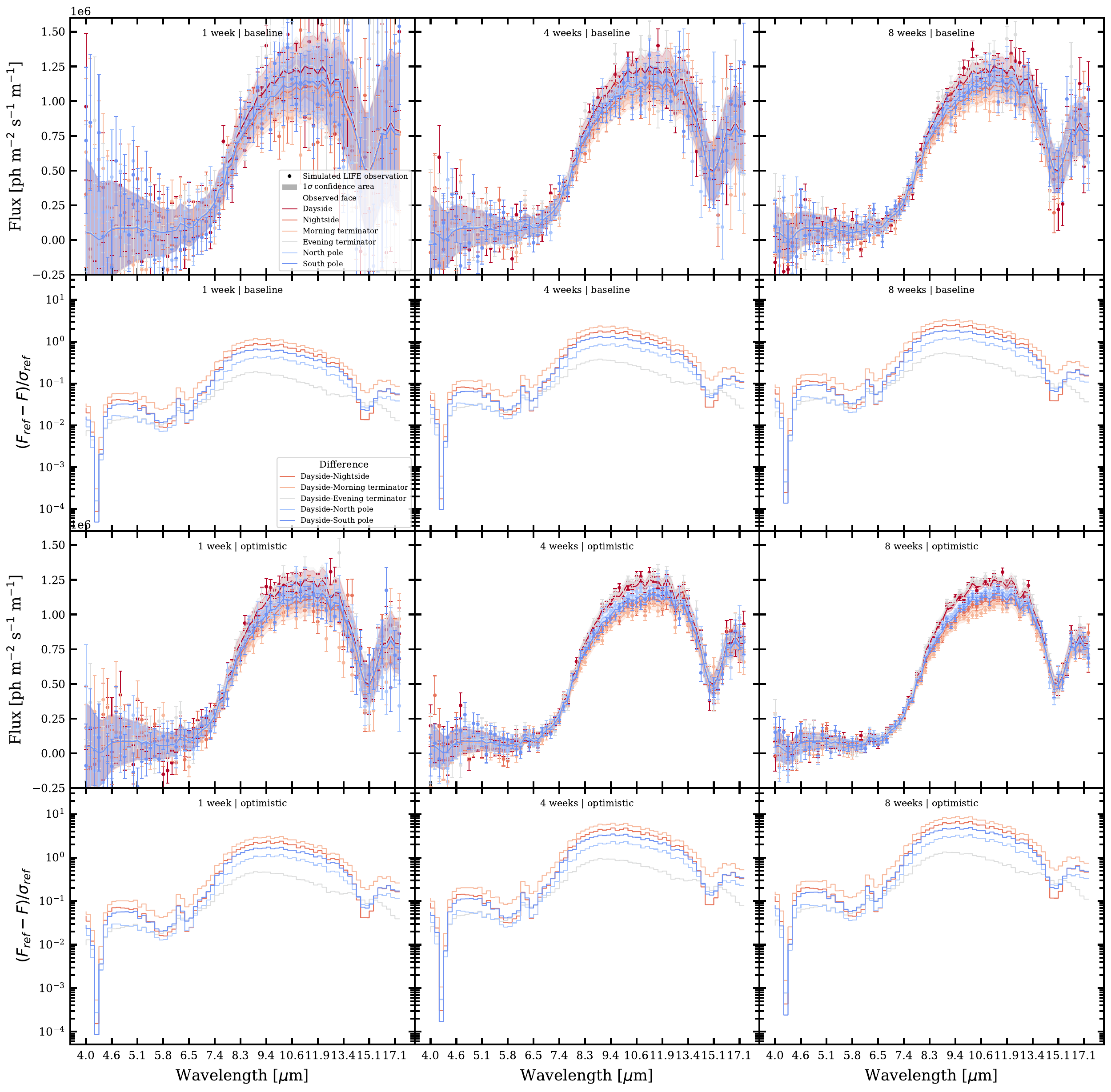} 
\caption{LIFEsim synthetic observation of Teegarden's Star b with a resolution of 50 and integration times of one week, one month, and two months assuming a baseline and an optimistic scenario. Rows 1 and 3 depict the thermal emission integrated over the six hemispheres of the planet. The shaded areas represent the 1$\sigma$ confidence level. Rows 2 and 4 show the statistical significance of the detected difference between the dayside and the other hemispheres.}
\label{fig:Mstar_LIFEsim}
\end{figure*}

Rows 1 and 3 of Figure \ref{fig:Mstar_LIFEsim} show the emission spectrum of Teegarden b for each observed hemisphere, assuming a baseline and an optimistic scenario respectively. The error bars correspond to an observation length of one week, one month, or two months. The shaded 1$\sigma$ confidence areas are defined within one standard deviation of the noise around the signal. The noise is assumed normally distributed. Rows 2 and 4 of the figure show the statistical significance of the detected difference between the dayside centered on the substellar point and the other hemispheres.
The baseline scenario results in a significance smaller than 1-sigma in most cases, however extending the observation to four weeks allows most hemispheres to be distinguished from the dayside at more than a 1$\sigma$ confidence level. Pushing the observation to two months allows all hemispheres to be differentiated from the dayside with more than a 1$\sigma$ confidence level, even exceeding 3$\sigma$ in some cases. The optimistic scenario improves LIFE's capabilities substantially in this case study. Exceeding a significance of 1$\sigma$ for most hemispheres can be achieved with a week-long observation. 

\begin{figure*}[htb]
\centering
\includegraphics[width=0.79\textwidth]{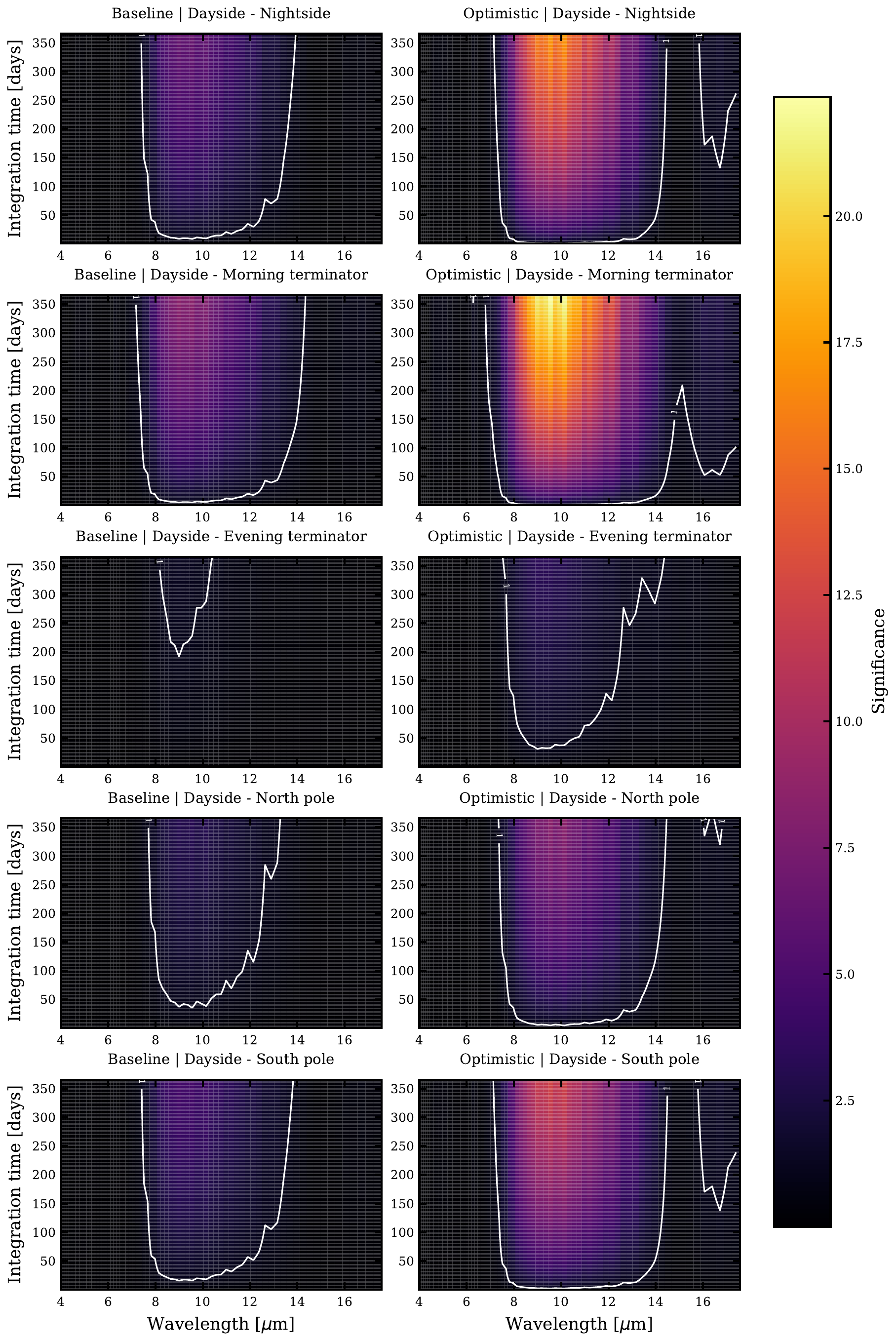} 
\caption{Integration time as a function of wavelength for Teegarden's Star b, assuming an instellation of 1481 Wm$^{-2}$ and either a baseline or an optimistic scenario. The dayside hemisphere is taken as the reference. White contours indicate a confidence level of 1$\sigma$.}
\label{fig:Mstar_LIFEsim2D}
\end{figure*}

Figure \ref{fig:Mstar_LIFEsim2D} explores the full range of integration times from one day to a year. It shows that given enough observation time, significant confidence levels can be achieved with LIFE, which is a possible sign of a further capability of the mission to achieve a spatial resolution of the planet of an extent to be explored in the future. 

A hypothetical planet orbiting a Sun-like star and receiving the same instellation as Teegarden's Star b would most likely not be tidally-locked, which means that the relevant difference to look at here is between low and high latitudes. Again, as shown by appendix figure \ref{fig:Gstar_rot_Faces}, the thermal emission pattern is mostly driven by the cloud cover. 
Despite receiving the same instellation, under the light of a G star the planet yields a smaller thermal emission by a factor of about 1.6 on average. The incoming starlight from the G star does not contain as much infrared radiation as the M star, so there is less atmospheric heating by absorption, which results in a cooler climate and a smaller outgoing longwave radiation \citep{eager2020implications}. This yields a smaller thermal emission amplitude compared to the tidally-locked M star case. The lack of tidal locking also homogenizes the emission across longitudes, yielding much closer signals for each hemisphere of the planet, which makes them much more challenging to disentangle.

\begin{figure}[htb]
\centering
\includegraphics[width=0.49\textwidth]{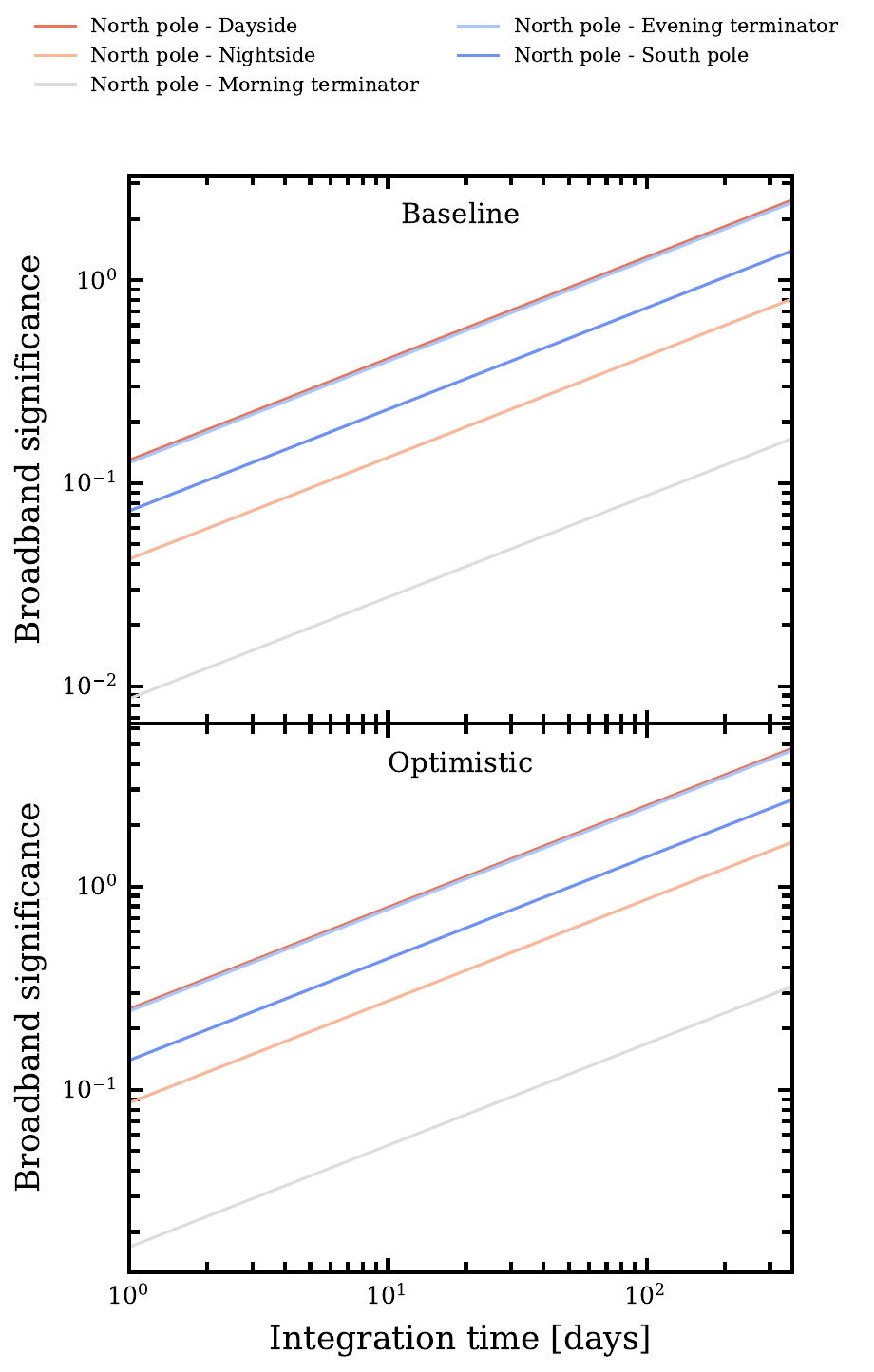} 
\caption{Broadband significance as a function of integration time for a planet orbiting a G star at a distance such that it receives 1481 Wm$^{-2}$, assuming a baseline and an optimistic scenario. The hemisphere centered on the north pole is taken as the reference.}
\label{fig:broadband_Gstar}
\end{figure}

Figure \ref{fig:broadband_Gstar} shows that although the optimistic scenario may allow the separation of the north pole hemisphere from the dayside with a 3$\sigma$ confidence level in about 50 days, overall the analysis would necessitate an unreasonable amount of time.

\section{Discussion} \label{sec:discussion}

Our results indicate that LIFE has realistic prospects for separating the thermal emission from different hemispheres of Teegarden b and other tidally-locked temperate exoplanets with similar orbital parameters. In broadband observations, this can be achieved within moderate observation times, and would allow the mission to probe whether any total day-night difference can be observed in the first place. If any are observed with high enough confidence levels, LIFE could invest longer observing times to extract more information from the emission spectra. Reaching significant confidence levels with spectrally-resolved observations may require at least a week-long observation even if LIFE achieves capabilities expected in the optimistic scenario. In addition, LIFE will be able to obtain thermal emission spectra along the orbit, giving it access to a continuum of hemispheres, as long as the orbit is not face-on, which is statistically unlikely. From all these spectra thermal maps of the planet may be reconstructed to retrieve spatial patterns of thermal emission. This would be a valuable indicator of day-night heat distribution, allowing retrievals of day-night temperature contrasts. For example, a strong day–night contrast suggests poor heat transport and a possible atmospheric collapse on the nightside, while modest contrasts argue for an efficient circulation which may be explainable by oceanic heat transport \citep{hu2014role}, although this argument may not work if the planet is too close to the inner edge of its habitable zone \citep{yang2019ocean}, which is a possibility suggested in the work done by \cite{boukrouche2025near}.

Such patterns would also potentially be linkable to moisture or cloud patterns, since these are primary drivers of the outgoing longwave radiation pattern. The appendix figure \ref{fig:CERES} illustrates this by showing Earth observations of the outgoing longwave radiation and the cloud area fraction averaged over the month of June 2025. Smaller polar and nightside emissions compared to the dayside, in the tidally locked case, can be explained by clouds formed from moisture carried from the dayside by advection. In temperate conditions, the presence of moisture most likely implies the presence of clouds and the existence of a hydrology. Greenhouse gases such as CO$_2$ would have a much more homogeneous effect on the OLR, and water clouds may be the most likely candidate. It would however be worth exploring possible false positives which may offer other explanations for these patterns.

Exploring different orbital geometries, which would probe the hemispheres differently, will be needed until we can constrain the geometry of the orbit of Teegarden's Star b. Additionally, it will be helpful to determine the resolutions achievable by reconstructing thermal maps based on spectra obtained along the orbit. Obtaining sub-hemispheric resolutions may be required. Even if this resolution ends up being very coarse, it might be the first instance of a temperate rocky planet potentially spatially resolvable.  

Teegarden's Star b will be an important testing bed for LIFE. If its orbit is not face-on, the postprocessing of the thermal emission spectra along its orbit may reveal crucial information related to its potential habitability, most notably the presence of moisture and clouds and the day-night temperature heat transport. All these pieces have the potential to robustly imply the presence of an ocean.

\clearpage
\acknowledgements
Some of the computations were enabled by resources provided by the National Academic Infrastructure for Supercomputing in Sweden (NAISS), partially funded by the Swedish Research Council through grant agreement no. 2022-06725. This work was also supported by an interdisciplinary postdoctoral fellowship issued by the Section for Mathematics and Physics at Stockholm University. We thank Dr. Felix Dannert for his help with LIFEsim and the reviewer for helping improve the manuscript.

\newpage
\textit{Software:} \textsc{numpy} \citep{numpy:2020}, \textsc{scipy} \citep{scipy:2001}, \textsc{matplotlib} \citep{matplotlib:2007}, \textsc{seaborn} \citep{seaborn:2018}, \textsc{socrates} \citep{edwards1996socrates}, \textsc{LIFEsim} \citep{dannert2022large}

\vspace{4cm}
\bibliography{references}{}
\bibliographystyle{aasjournal}

\appendix
\renewcommand{\thefigure}{A.\arabic{figure}} 
\setcounter{figure}{0} 

\begin{figure*}[htb]
\centering
\includegraphics[width=0.79\textwidth]{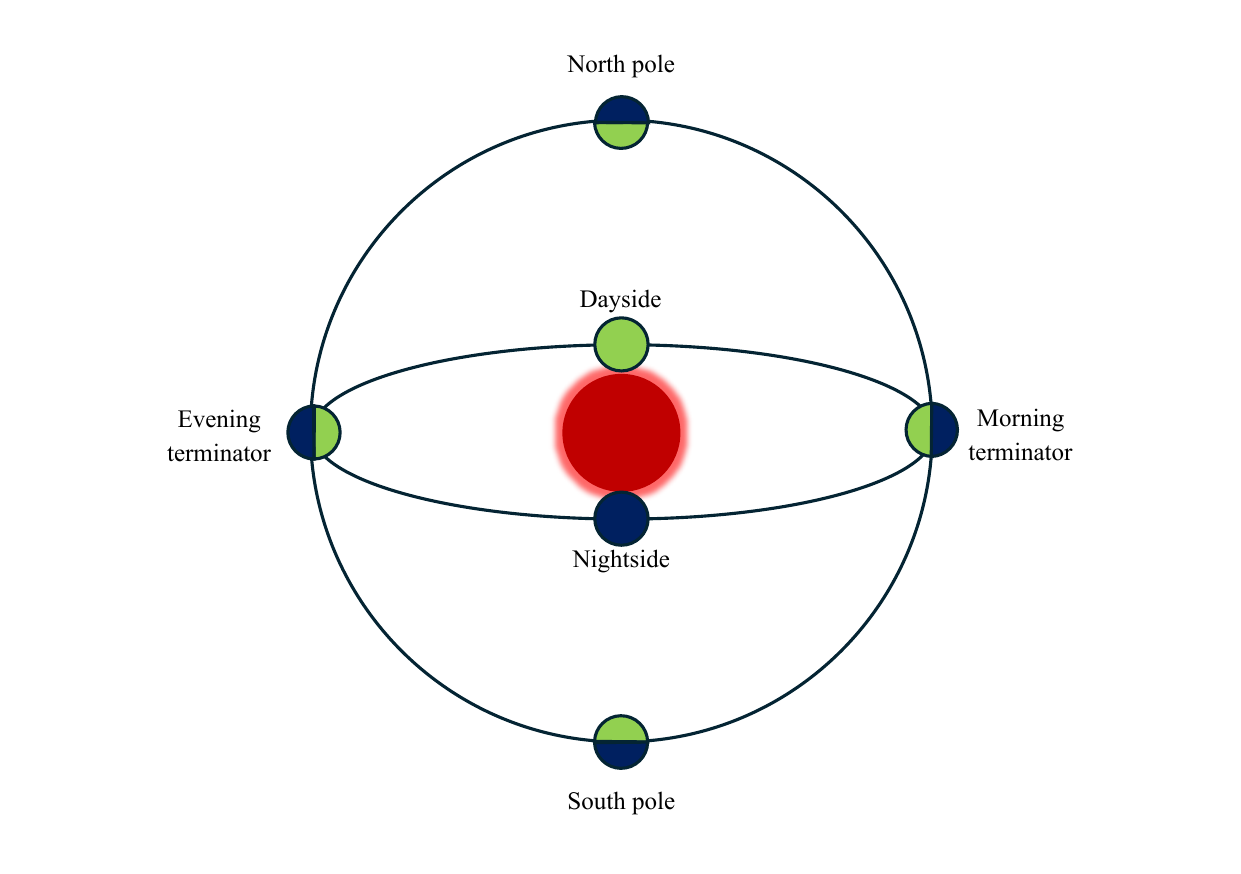} 
\caption{Illustration of the six hemispheres.}
\label{fig:A1_orbital_cartoon}
\end{figure*}

Figure \ref{fig:A1_orbital_cartoon} illustrates the location of the planet along its orbital phase for the six visible hemispheres considered in this work.

\begin{figure}[ht]
    \centering
    \begin{minipage}{0.45\textwidth}
        \centering
        \includegraphics[width=\linewidth]{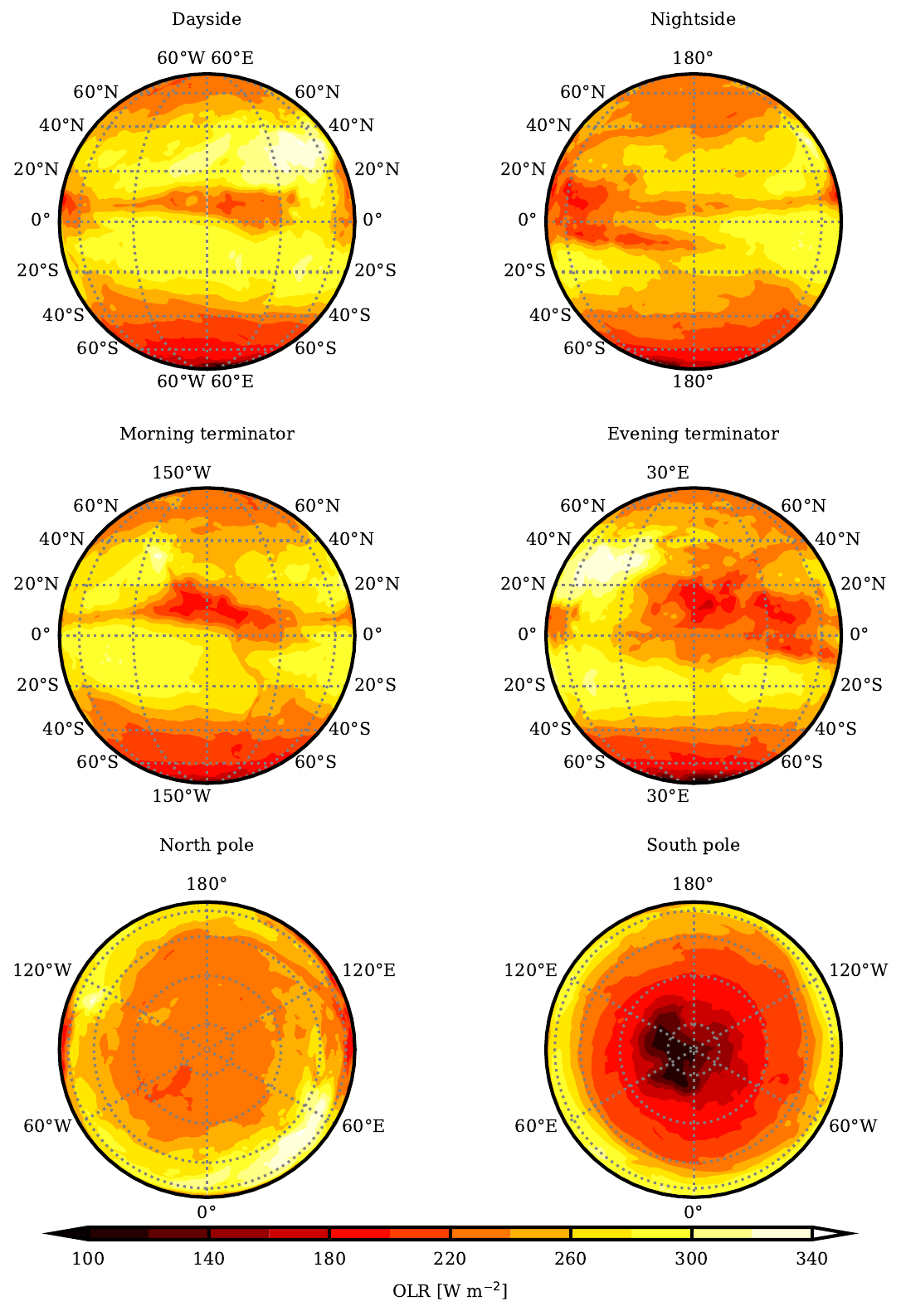}
        \caption{}
    \end{minipage}\hfill
    \begin{minipage}{0.45\textwidth}
        \centering
        \includegraphics[width=\linewidth]{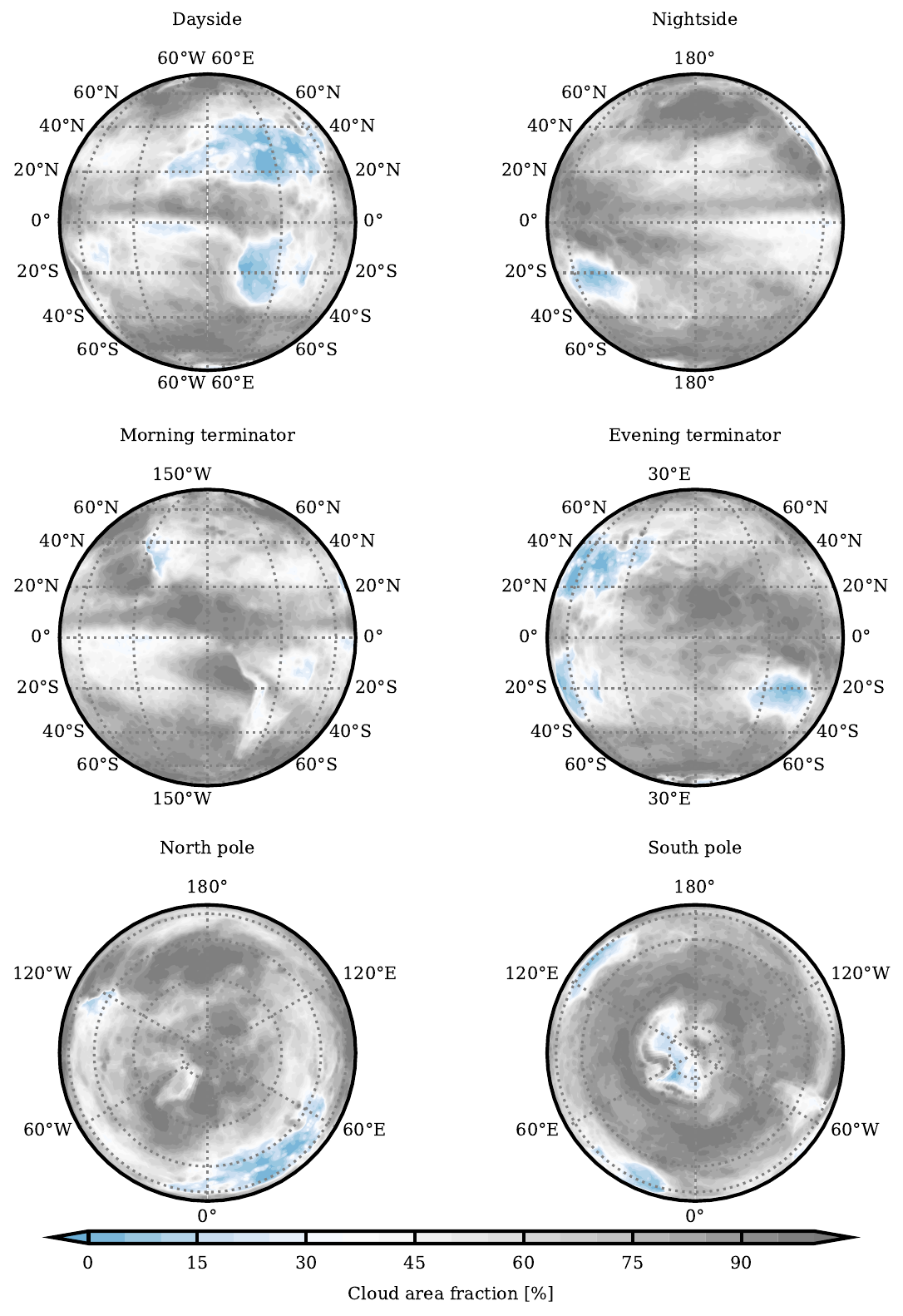}
        \caption{}
    \end{minipage}
    \label{fig:CERES}
\end{figure}

Figure \ref{fig:CERES} shows the observed outgoing longwave radiation and cloud area fraction on Earth averaged over the month of June 2025. This data was obtained from the edition 4.2.1 Energy Balanced and Filled (EBAF) data product the Clouds and the Earth’s Radiant Energy System (CERES) project \citep{wielicki1996clouds,kato2025seamless}.

\begin{figure}[htbp]
    \centering
    \begin{minipage}[t]{0.45\textwidth}
        \includegraphics[width=\linewidth]{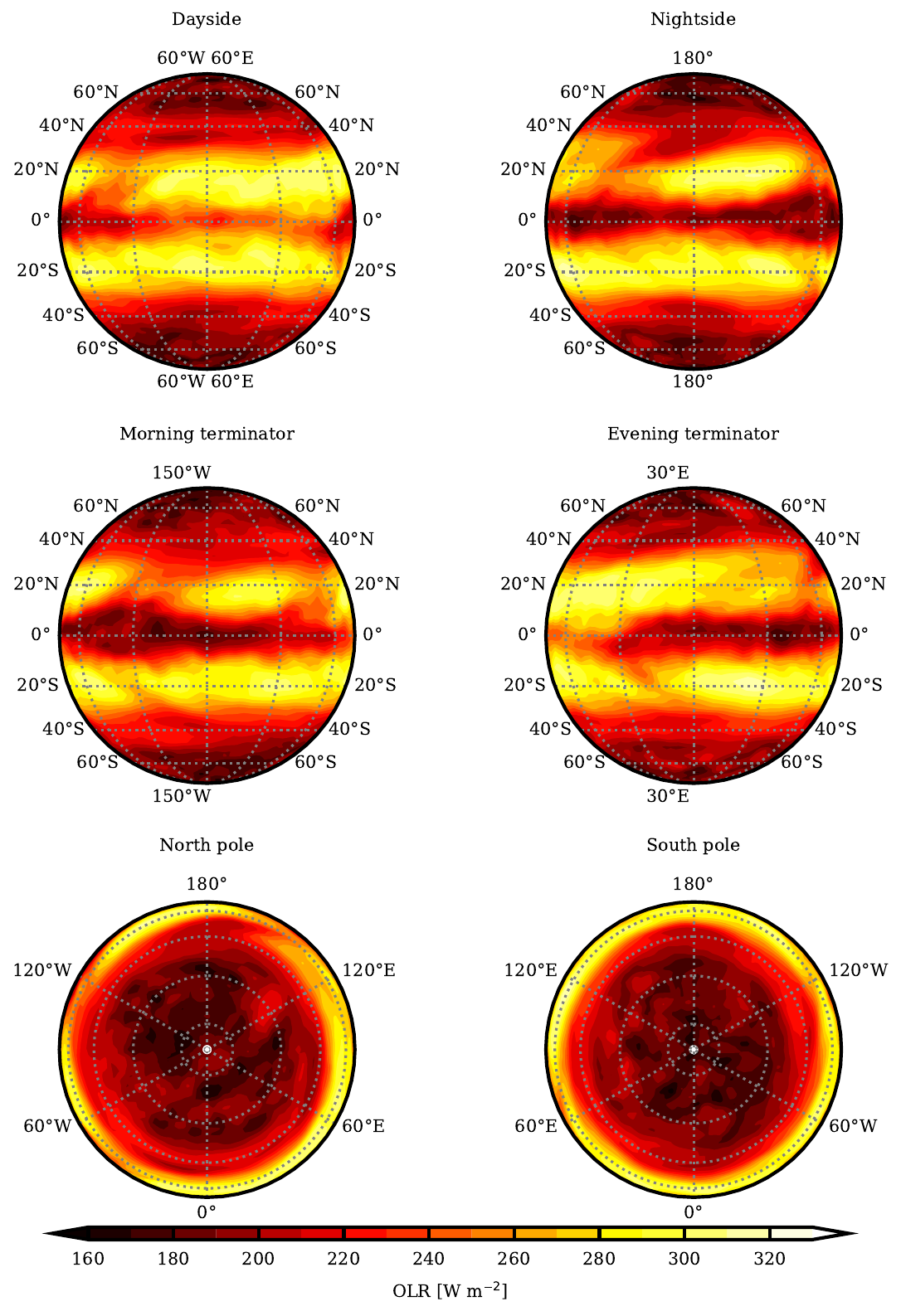}
    \end{minipage}\hfill
    \begin{minipage}[t]{0.45\textwidth}
        \includegraphics[width=\linewidth]{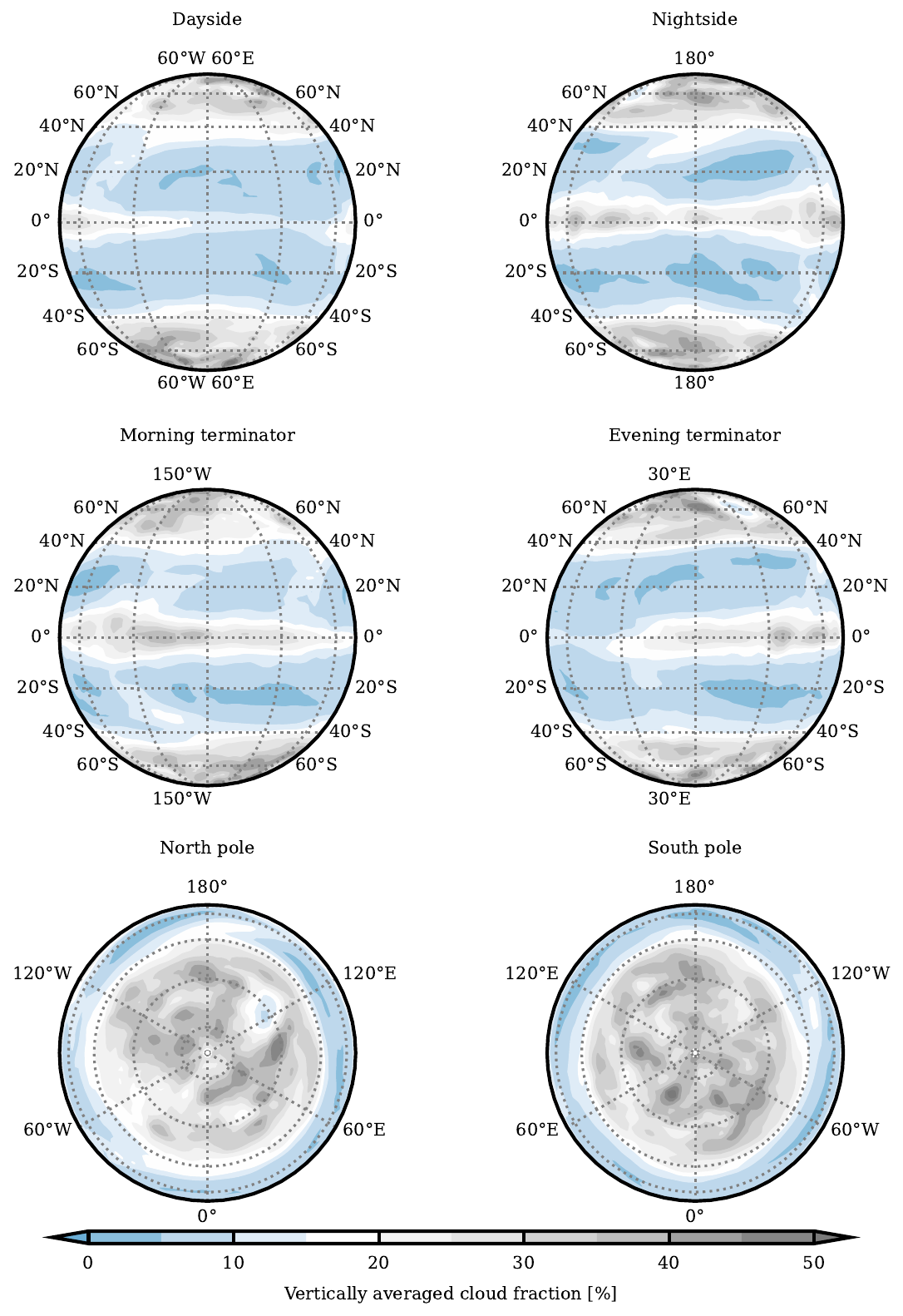}
    \end{minipage}
    \caption{Outgoing longwave radiation and vertically averaged cloud fraction around a Sun-like star with an Earth-like rotation rate and at a distance such that the planet receives the same instellation as Teegarden's Star b.}
    \label{fig:Gstar_rot_Faces}
\end{figure}

Figure \ref{fig:Gstar_rot_Faces} show the outgoing longwave radiation and the vertically-averaged cloud cover around a Sun-like star with an Earth-like rotation rate and at a distance such that the planet receives 1481 Wm$^{-2}$, as on Teegarden's Star b according to \cite{dreizler2024teegarden}.

%
%

%
%

\begin{figure*}[htb]
\centering
\includegraphics[width=0.79\textwidth]{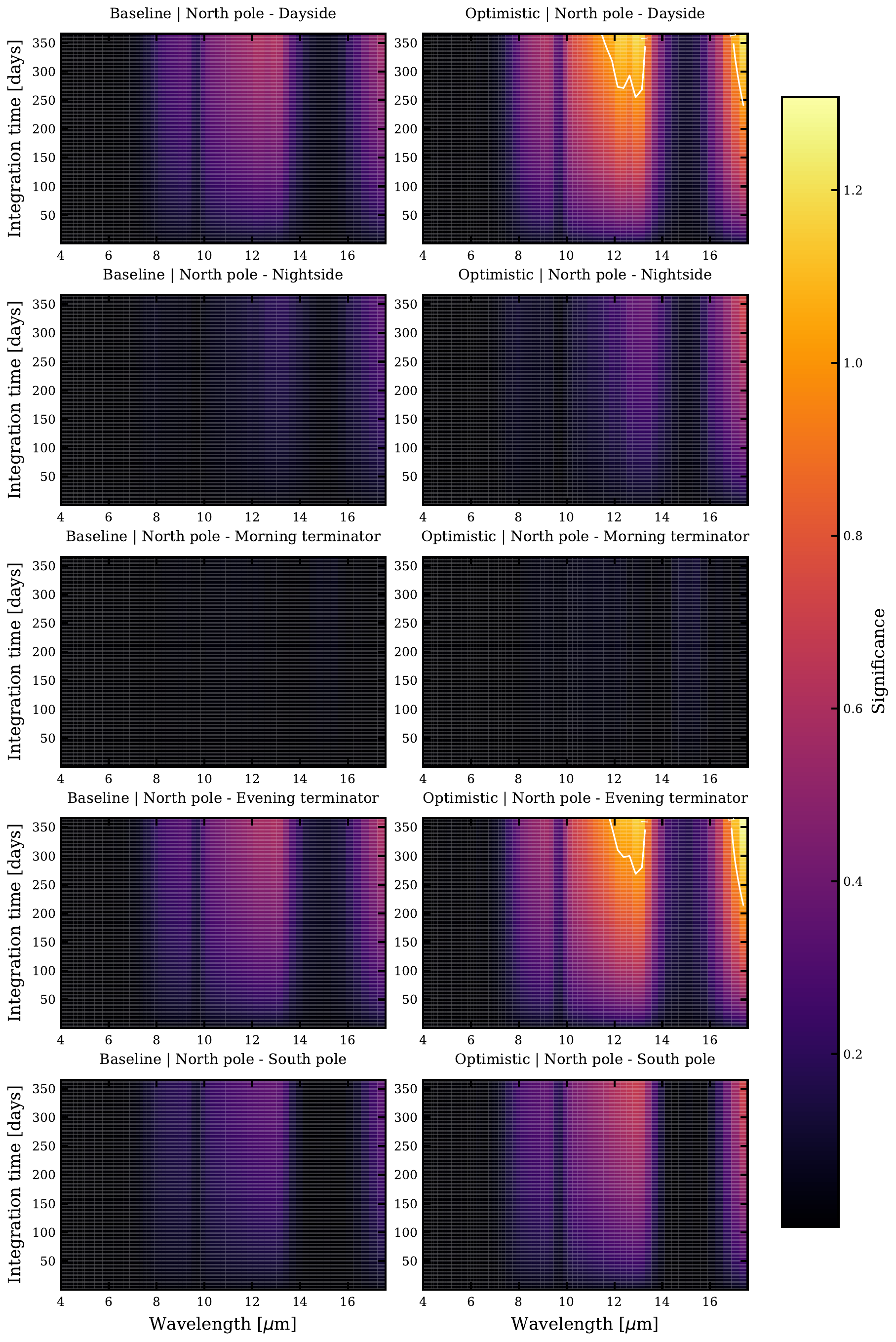} 
\caption{Integration time as a function of wavelength for a planet orbiting a G star at a distance such that it receives 1481 Wm$^{-2}$, assuming a baseline and an optimistic scenario. The hemisphere centered on the north pole is taken as the reference.}
\label{fig:Gstar_LIFEsim2D}
\end{figure*}

Figure \ref{fig:Gstar_LIFEsim2D} shows the results of the sweep over the integration time for each hemispheric comparison in the case of the planet around a G-class star. It takes the hemisphere centered on the north pole as reference, since we expect equator-to pole differences in thermal emission to be larger in this case. However, only two out of five comparison cases yield a confidence level above 1$\sigma$, and even then only when the optimistic scenario is assumed, and for integration times longer than 250 days.

%

\end{document}